\documentclass[letterpaper]{article} 
\usepackage{aaai23}  
\usepackage{times}  
\usepackage{helvet}  
\usepackage{courier}  
\usepackage[hyphens]{url}  
\usepackage{graphicx} 
\usepackage{natbib}  
\usepackage{caption} 
\usepackage{algorithm}
\usepackage{algorithmic}

\usepackage{newfloat}
\usepackage{listings}
\DeclareCaptionStyle{ruled}{labelfont=normalfont,labelsep=colon,strut=off} 
\floatstyle{ruled}
\newfloat{listing}{tb}{lst}{}
\floatname{listing}{Listing}
\usepackage{amsmath,amsfonts}
\usepackage{graphicx}
\usepackage{enumitem}
\usepackage{comment}

\newcommand{\sets}[1]{\ensuremath{\mathcal{#1}}}

\usepackage{xcolor}

\usepackage{amsthm}

\theoremstyle{plain}
\newtheorem{definition}{Definition}[section]

\newtheorem{proposition}{Proposition}

\AtEndDocument{\refstepcounter{proposition}\label{finalthm}}

\title{Fairness in Contextual Resource Allocation Systems:\\ Metrics and Incompatibility Results}
\author {
    \equalcontrib Nathanael Jo\textsuperscript{\rm 1,\rm 2}, 
    \equalcontrib Bill Tang\textsuperscript{\rm 1}, 
    Kathryn Dullerud\textsuperscript{\rm 1},  
    Sina Aghaei\textsuperscript{\rm 1},
    Eric Rice\textsuperscript{\rm 1},
    Phebe Vayanos\textsuperscript{\rm 1}
}
\affiliations {
    \textsuperscript{\rm 1} USC Center for AI in Society, Los Angeles, CA \\
    \textsuperscript{\rm 2} Stanford University, Palo Alto, CA \\
    nathanjo@stanford.edu, \{yongpeng, kdulleru, saghaei, ericr, phebe.vayanos\}@usc.edu
}

\usepackage{bibentry}

\begin{document}
\maketitle

\begin{abstract}
We study critical systems that allocate scarce resources to satisfy basic needs, such as homeless services that provide housing. These systems often support communities disproportionately affected by systemic racial, gender, or other injustices, so it is crucial to design these systems with fairness considerations in mind. To address this problem, we propose a framework for evaluating fairness in contextual resource allocation systems that is inspired by fairness metrics in machine learning. This framework can be applied to evaluate the fairness properties of a historical policy, as well as to impose constraints in the design of new (counterfactual) allocation policies. Our work culminates with a set of incompatibility results that investigate the interplay between the different fairness metrics we propose. Notably, we demonstrate that: \emph{1)} fairness in allocation and fairness in outcomes are usually incompatible; \emph{2)} policies that prioritize based on a vulnerability score will usually result in unequal outcomes across groups, even if the score is perfectly calibrated; \emph{3)} policies using contextual information beyond what is needed to characterize baseline risk and treatment effects can be fairer in their outcomes than those using just baseline risk and treatment effects; and \emph{4)}~policies using group status in addition to baseline risk and treatment effects are as fair as possible given all available information. Our framework can help guide the discussion among stakeholders in deciding which fairness metrics to impose when allocating scarce resources.

\end{abstract}

\section{Introduction}\label{sec:introduction}

Many of our social and health service systems that operate in high-stakes settings are severely under-resourced. These systems allocate scarce resources that satisfy basic human needs, such as hospitals that provide treatments, social services that offer preventive care, or homeless services that provide shelter and housing. Inadequate allocation in these systems can have grave consequences; for instance, a social service program may exclude those most at risk of suicide from an appropriate preventive intervention.

Since these systems often support communities disproportionately affected by systemic racial, gender, or other injustices, they must be sensitive to the worsening of preexisting inequalities. As such, the fairness of a high-stakes resource allocation system must be easily scrutinized and understood by key stakeholders, which include policymakers and community members the system is designed to help. Indeed, even the perception of bias or lack of transparency can result in a loss of trust and participation. Systems that do not align with the values of their stakeholders may even get defunded, further exacerbating resource scarcity.

Our goals in this paper are a) to establish a simple framework for stakeholders to evaluate the fairness of such high-stakes resource allocation systems, and b) to guide policymakers in choosing which fairness requirements to impose in the design of new (counterfactual) allocation policies. In these systems, a central decision maker makes recommendations for how resources should be allocated. Such policies rely on \emph{contextual} information, i.e., they are based on an individual's intrinsic characteristics (or covariates) which are validated by case managers, as opposed to an individual's preferences (as in the mechanism design literature). Therefore, our learning problem closely resembles that of the classical machine learning framework. Instead of predicting outcomes from features, however, we additionally consider the causal effect of resources (or treatments) on covariates. 

We first present a menu of options for evaluating fairness at different stages of the allocation process that are inspired by popular fairness notions in the machine learning literature. We then present incompatibility results between these notions that illuminate the concrete decisions that policymakers need to make in deciding how to allocate resources. For instance, is it acceptable to use protected characteristics like race in deciding which resource an individual should receive? Is it sufficient to only use a vulnerability score to prioritize individuals for resources, or should one necessarily include more context (e.g., information on treatment effects)?


Our results in this paper apply to many high-stakes resource allocation systems. To streamline exposition, we focus on one concrete application, that of allocating scarce housing resources to people experiencing homelessness.

\textbf{Housing Allocation for People Experiencing Homelessness in Los Angeles.} Many cities face dramatic shortages of housing resources used to support those experiencing homelessness. In Los Angeles County, for example, there are over 69,000 homeless persons and only around 30,000 permanent housing units as of the latest counts in 2022 \cite{lahsa_pit_ind, lahsa_hic}. Communities have attempted to address this problem by creating Coordinated Entry Systems (CES) where agencies pool their housing resources in a centralized system called a Continuum of Care (CoC). Agencies then distribute housing and services funding for people experiencing homelessness from this pooled resource. In Los Angeles, the CoC is directed by the Los Angeles Homeless Services Authority (LAHSA), who conducts the following process for allocating housing resources:





\begin{enumerate}
    \item \textbf{Assessment} -- Persons seeking housing are first assessed for eligibility and vulnerability by filling out a self-reported survey known as the Vulnerability Index - Service Prioritization Decision Assistance Tool (VI-SPDAT), which gives a score from 0 to 17. A higher score indicates higher vulnerability.
    \item \textbf{Enrollment} -- Case managers then use this score  to ``enroll'' the individual into a housing resource, which may be one of the following interventions: Permanent Supportive Housing (PSH) or Rapid Re-Housing (RRH), which are intended for individuals with severe and moderate housing needs, respectively.
    \item \textbf{Allocation} -- Finally, individuals may receive a resource. Note the important distinction between what the individual has been \textit{enrolled} in and what they have been \textit{given}. Policymakers can only recommend connections to particular resources; the actual allocation depends on a variety of factors, including the availability of the resource, whether or not a landlord is willing to accept certain individual to their housing community, and many more.  
    \item \textbf{Outcome} -- The impact of that resource is evaluated. While this step is not explicitly part of the matching process, we include it in order to be able to quantify the performance of a housing allocation policy. For example, the outcome being tracked may correspond to whether or not a homeless person returns to homelessness within~24 months of receiving a resource.
\end{enumerate}

In recent years, the fairness of CES has been put to question. In 2018, the Los Angeles Homeless Services Authority's Ad Hoc Committee on Black People Experiencing Homeless published their report mentioning that while CES seems to match Black individuals to housing resources at equal rates to their White counterparts, they still ``experience a higher rate of returns to homelessness than all other race and ethnic groups''~\cite{lahsa2018report}. Our work aims to contextualize these claims by providing various metrics to evaluate the fairness of a resource allocation system that are more nuanced than the ones practitioners traditionally use.

\subsection{Contributions}\label{section:contributions}
Our key contributions are:

\begin{itemize}
    \item We propose a unifying framework for measuring fairness in high-stakes resource allocation systems which has several advantages: it is intuitive, in that any decision maker with little to no knowledge of algorithmic fairness can understand the fairness concepts; it is adaptable, in that our framework allows for the evaluation of arbitrary counterfactual policies; it is also comprehensive in considering the various sources of discrimination that could be present in a complex allocation system. Throughout this paper, we present our fairness framework in the general sense but will specifically focus on the motivating example of allocating resources for people experiencing homelessness in Los Angeles.
    
    \item We present incompatibility results showing that\: \emph{1)} fairness in allocation and fairness in outcomes are usually incompatible; \emph{2)} policies that prioritize based on a vulnerability score will usually result in unequal outcomes across groups, even if the score is perfectly calibrated; \emph{3)} policies using contextual information beyond what is needed to characterize baseline risk and treatment effects can be fairer in their outcomes than those using just baseline risk and treatment effects; and \emph{4)} policies using group status in addition to baseline risk and treatment effects are as fair as possible given all available information.
    

\end{itemize}

The remainder of this paper is organized as follows. The rest of Section~\ref{sec:introduction} positions our paper within the related literature. In Section~\ref{section:statistical_fainess}, we formalize our proposed notions of group fairness. Section~\ref{section:incompatibility_results} presents our incompatibility results when using a combination of these fairness notions together. Finally, we discuss the practical implications of our incompatibility results in Section~\ref{section:discussion}.

\subsection{Literature Review}
Our work is closely related to the fields of fairness in machine learning and resource allocation. We now position our work in these two literature streams.

\subsubsection{Fairness in Machine Learning.}


The literature on algorithmic fairness is extensive. Of particular relevance to our work is enforcing ``group fairness'', i.e., \textit{statistical fairness} over segments of the population. This is in contrast to ``individual'' fairness, which requires that individuals with similar covariates be classified in the same way \cite{Dwork2012}. We focus on group fairness notions because they are intuitive and are more aligned with how practitioners evaluate fairness. There are a number of fairness metrics that have been proposed in the literature. For instance, \textit{statistical parity} enforces that the probability of receiving a positive class is equal across all sensitive groups~\cite{Dwork2012}. As an extension to \textit{statistical parity}, \textit{conditional statistical parity} was introduced by~\citet{Corbett2017}, which stipulates that a classifier should assign a positive class at equal rates across all protected groups, conditional on some legitimate feature(s) that affect the outcome.

Other fairness notions incorporate the quality of a classifier's decisions, such as \textit{equalized odds} where all protected groups should have the same true positive rates (TPR) and false positive rates (FPR)~\cite{Hardt2016}. A simpler version of \textit{equalized odds} is \textit{equality of opportunity}, where only equal TPR is enforced. Approaches that include these two notions include~\citet{Hardt2016, Zafar2017b, agarwal2018reductions}.

There are many other notions of fairness such as \textit{error rate balance}~\cite{Feldman2015}, \textit{predictive parity}~\cite{Chouldechova2017}, \textit{well-calibration}~\cite{crowson2016assessing,flores2016false}, and \textit{balance of positive (negative) class}~\cite{Kleinberg2016}. We refer the interested reader to~\citet{barocas-hardt-narayanan, Mehrabi2021} for in-depth reviews of the literature.

An important line of work in fair machine learning investigates incompatibility/impossibility results among different notions of fairness. For instance, \citet{Kleinberg2016} show the incompatibility of \textit{well-calibration} and \textit{balance of positive/negative class} and \citet{Chouldechova2017} shows that \textit{well-calibration} and \textit{error rate balance} are incompatible except in trivial cases. 

Another relevant branch of research in fair machine learning is the study of how ``race-blind'' practices might actually hinder fairness. \citet{huq2018racial} provides a perspective on how including group membership could work in a legal setting and \citet{kleinberg2018algorithmic} show that race-aware predictors outperform race-blind predictors in terms of fairness. Our work adds another dimension to this literature by expanding this stream of work into the resource allocation setting.

\subsubsection{Fairness in Resource Allocation.} 
Within the literature on fairness in resource allocation, works that are most closely related to ours are those that have attempted to bridge the gap between fair machine learning and fair resource allocation/mechanism design~\cite{Finocchiaro2021, das2021local}. However, these papers largely provide guiding frameworks for how problems connecting machine learning and resource allocation should be approached rather than defining fairness notions that bridge the two. 
There are several papers that approach the problem of fair resource allocation through a central decision maker. \citet{rodolfa2020case} use the distribution of recall across protected groups to improve fairness, while also adjusting for how their measure of fairness might affect long-term outcomes. \citet{rodolfa2021} build on the previous work by showing they do not sacrifice accuracy in applying their notion of fairness. \citet{Elzayn2019} present a notion of allocative fairness inspired by equality of opportunity in the classification setting, and \citet{Donahue2020} extend this work by providing an upper bound on the price of fairness. However, these papers focus on just one notion of fairness and thus none examine possible incompatibilities between different fairness notions.

Another line of work in this literature focuses on designing policies that satisfy fairness notions through linear constraints. Some applications include kidney allocation \cite{Bertsimas2013} and allocation of housing resources \cite{Azizi2018, rahmattalabi2022learning}. Some of the fairness notions imposed in these works overlap with the ones we propose in our framework. However, none of these papers study the interplay between different notions of fairness. \citet{Vayanos2021} propose a preference elicitation algorithm and show how it can be used to elicit the preferences of policymakers over various notions of fairness and efficiency. Yet, they do not present any incompatibility results between these fairness notions. There are also other works that compare fairness notions in non-contextual resource allocation problems such as facility location~\cite{kumar2000fairness}, scheduling~\cite{kumar2000fairness, baruah1993proportionate}, and bandwidth assignments in networks~\cite{kumar2000fairness,  lan2010axiomatic}.



There are a number of papers that focus on fairness in mechanism design. However, they focus on the utility derived from allocating a resource to an individual. For example, \citet{freeman2020best} develop algorithms for constructing allocation policies that are simultaneously exactly fair ex-ante and approximately fair ex-post and compare fairness notions such as envy-freeness, group  fairness, and proportionality. \citet{mashiat2022trade} outline several notions of fairness related to equalizing average regret and improvement in allocation across groups. \citet{noagentbehind} and \citet{ Benabbou2019} focus on the incompatibility of fairness and efficiency in terms of envy-freeness and pareto-optimality. Other works model the allocation system using self-interested agents trying to obtain a good(s) through a market~\cite{Conitzer2019, scarlett2021one, Benabbou2019, Benabbou2019b, Barman2018, Fain2018, Raman2021}. Our work focuses specifically on the interplay of fairness notions at different phases of allocation and on contextual resource allocation systems.

\section{Statistical Fairness}\label{section:statistical_fainess}
We consider a system that allocates scarce resources of different (finite) types, which we index in the set $\sets{T}$ (e.g., PSH or RRH). These resources must be allocated to individuals who are characterized by their feature vector $X \in \sets{X} \subseteq \mathbb{R}^n$, which includes their sensitive attribute(s) $G \in \sets{G}$, where $\sets{G}$ comprises the levels of (potentially intersection of multiple) sensitive attribute(s) (e.g., race or sex). Each individual is also characterized by their potential outcomes $Y^t \in \sets{Y} \subseteq \mathbb{R}$ under each resource $t \in \sets{T}$. These potential outcomes could be interpreted as one's utility given treatment (e.g., long-term health or return to homelessness), though they are not a function of an individual's preferences. The system uses a (possibly randomized) recommendation policy to assign each individual a treatment, $T = \pi(X) \in \sets{T}$, i.e., $\pi: \sets{X} \mapsto \sets{T}$. This recommendation may not be the same as the treatment that the individual is given, denoted by $A \in \sets{T}$, which is determined by a (possibly randomized) allocation rule $\mu: \sets{X} \mapsto \sets{T}$. While we have defined our policies $\pi$ and $\mu$ in the general case for $X$, which includes $G$, in many sensitive settings using protected characteristics is prohibited to prevent discrimination. Therefore, we also define $\mathcal{X}_{-G}$ as the set of features excluding group status $G$. Finally, for a given $\pi, \mu$, we assume an unknown joint distribution $\mathbb{P}$ over $(X, T, A, \{Y^t\}_{t\in \mathcal{T}})$. Note that the random variable $\pi(x)$ is specified by the conditional probabilities $\mathbb{P}(\pi(x)=t|X=x)$ that each individual is assigned treatment $t$ conditional on covariates $x$ and $\mu(x)$ is specified by the conditional probabilities $\mathbb{P}(\mu(x)=t|X=x)$ that each individual receives treatment $t$ conditional on covariates $x$. Furthermore, deterministic policies are a special case of the randomized policies we have defined where the conditional probabilities $\mathbb{P}(\pi(x)=t|X=x)$ and $\mathbb{P}(\mu(x)=t|X=x)$ equal~1 for a single treatment and~0 for all other treatments. The goal, then, is to evaluate the fairness of the system with respect to different groups. 

\subsection{Enrollment Fairness}\label{sec:enrollment_fairness}

We now outline several fairness considerations during the enrollment stage, i.e., the results following a recommendation rule $\pi$.

\begin{definition}[Statistical Parity in Enrollments]\label{def:statistical_parity_enroll}

Statistical parity in enrollments is satisfied if the probabilities of enrolling in any given treatment are equal across sensitive groups, i.e., 
\begin{align*}\mathbb{P}(\pi(X)=t|G=g) = \mathbb{P}(\pi(X)=t|G=g') \\ \forall g, g' \in \sets{G}, t \in \sets{T}.
\end{align*}

\end{definition}

As mentioned in previous sections, statistical parity is perhaps one of the most well-known fairness notions in the machine learning literature. It is also used in evaluating many real-world systems, see e.g., LAHSA's Ad Hoc Committee on Black People Experiencing Homelessness report~\cite{lahsa2018report}. While it is simple and intuitive, this notion may not sufficiently protect groups that are more vulnerable at baseline. For instance, recall that homeless individuals in Los Angeles are assigned a risk score from 0 to 17 based on their vulnerability. Let us assume that we are enforcing statistical parity on race as the sensitive group. Since we do not condition on risk score, it is possible that an allocation policy may disproportionately assign treatments to those with lower vulnerability within each racial group -- which is the exact opposite of what we want to do. We therefore introduce the following definition.

\begin{definition}[Conditional Statistical Parity in Enrollments]\label{def:conditional_statistical_parity_enroll} Conditional statistical parity in enrollments is satisfied if the probabilities of enrolling in any given treatment are equal across sensitive groups, conditional on some ``legitimate feature(s)''. We define $L = f(X)$ as the random variable taken from the set of legitimate features $\sets{L}$, where $f: \mathbb{R}^n \mapsto \sets{L}$, i.e.,
\begin{align*}
\mathbb{P}(\pi(X)=t|G=g, L=\ell) = \mathbb{P}(\pi(X)=t|G=g', L=\ell)  \\ \forall g, g' \in \sets{G}, t \in \sets{T}, \ell \in \sets{L}.
\end{align*}
\end{definition}
In our motivating example, $L$, which is a function of covariates collected from a survey assessment, is the risk score from 0 to 17. Policies that assign or recommend based on a risk score are used in many applications (see \citet{optriskscore} for examples within medicine, criminal justice, and finance). Beyond conditioning on vulnerability, we also propose considering features measuring treatment effects as legitimate features since policymakers may be interested in targeting treatment effects to achieve outcome related fairness, which is detailed in Section \ref{subsec: outcomefairness def}.
\subsection{Allocative Fairness}

After being enrolled in a treatment, an individual may not actually receive their recommended treatment. We therefore define separate fairness notions for the allocation stage. While most of the definitions are similar to that of the enrollment stage, we include one more definition that is inspired by equalized odds. This is motivated by the need to ensure that the discrepancy between enrollment and allocation is roughly equal among all groups.

\begin{definition}[Equalized Faithfulness in Allocation]\label{def:equalized_odds_alloc} Equalized faithfulness in allocation is satisfied if the probabilities of receiving treatments other than the one initially recommended are equal across groups, i.e., 
\begin{align*}
\mathbb{P}(\mu(X)=t'|G=g, \pi(X)=t) = \\ \mathbb{P}(\mu(X)=t'|G=g', \pi(X)=t)  \\ \forall g, g' \in \sets{G}, t, t' \in \sets{T}.
\end{align*}
\end{definition}

Equalized faithfulness in allocation may be a useful mechanism when there are unaccounted biases towards certain groups between the enrollment and allocation stages. In our motivating problem of homelessness in Los Angeles, for example, some resources require that a housing complex's landlord approves of the person experiencing homelessness. Unfortunately, caseworkers have shared numerous cases of Black individuals being recommended a resource but not receiving it because no landlord is willing to accept them \cite{cplPSHinequity}.

Simply requiring equalized faithfulness between groups may lead to undesirable results when there are distributional differences in vulnerability between groups. For example, suppose group~$g$ is generally more vulnerable than group~$g'$. Imposing equalized faithfulness may result in an allocation where individuals from $g$ receive their recommended resource at much lower rates than their counterparts in $g'$ with the same vulnerability. This is possible so long as faithfulness is \textit{overall} the same across the two groups. A more reasonable policy might instead be more likely to allocate the recommended resource to more vulnerable individuals. We therefore introduce the notion of \textit{conditional} equalized faithfulness, which accounts for vulnerability.

\begin{definition}[Conditional Equalized Faithfulness in Allocation]\label{def:conditional_equalized_odds_alloc} Conditional equalized faithfulness in allocation is satisfied if the probabilities of receiving treatments other than the one initially recommended are equal across groups, conditional on some ``legitimate feature(s)'', i.e.,
\begin{align*}
\mathbb{P}(\mu(X)=t'|G=g, \pi(X)=t, L=\ell) = \\ \mathbb{P}(\mu(X)=t'|G=g', \pi(X)=t, L=\ell)  \\ \forall g, g' \in \sets{G}, t, t' \in \sets{T}, \ell \in \sets{L}.
\end{align*}
\end{definition}

We may also alter the definitions in Section~\ref{sec:enrollment_fairness} to get similar notions in the allocation setting.

\begin{definition}[Statistical Parity in Allocation]\label{def:statistical_parity_alloc} 
Statistical parity in allocation is satisfied if the probabilities of allocating any given treatment are equal across groups, i.e.,
\begin{align*}
\mathbb{P}(\mu(X)=t|G=g) = \mathbb{P}(\mu(X)=t|G=g')  \\ \forall g, g' \in \sets{G}, g \neq g', t \in \sets{T}.
\end{align*}
\end{definition}

\begin{definition}[Conditional Statistical Parity in Allocation]\label{def:conditional_statistical_parity_alloc} Conditional statistical parity in allocation is satisfied if the probabilities of allocating any given treatment are equal across groups, conditional on some \textit{legitimate} feature(s), i.e.,
\begin{align*}
\mathbb{P}(\mu(X)=t|G=g, L=\ell) = \mathbb{P}(\mu(X)=t|G=g', L=\ell)  \\ \forall g, g' \in \sets{G}, t \in \sets{T}, \ell \in \sets{L}.
\end{align*}
\end{definition}

While fairness in enrollment is something that policymakers control (because they necessarily assign the interventions), allocative fairness may not be possible to control since the actual allocation might depend on external factors like the availability of resources or discrimination (e.g., landlords might reject certain individuals from their housing communities). However, one may be able to affect allocative fairness by changing enrollment fairness assuming that faithfulness in allocation remains unchanged. Despite these challenges, allocative fairness is nonetheless important to measure because it reflects which resources individuals truly receive. We thus introduce these definitions separately to highlight the distinct challenges faced in the allocation stage versus the enrollment stage.

\subsection{Outcome Fairness}\label{subsec: outcomefairness def}
Lastly, we may impose similar notions of fairness from the outcome perspective, where outcomes could be some personal utility upon receiving treatment (e.g., a return to homelessness or health outcome). These definitions are useful in situations where imposing fairness in allocation is inappropriate because they exacerbate existing inequalities. For instance, Black individuals experiencing homelessness in Los Angeles may be assigned housing resources at equal rates to their White counterparts, but they still ``experience a higher rate of returns to homelessness than all other race and ethnic groups''~\cite{lahsa2018report}. 

For simplicity, we introduce these fairness notions assuming that potential outcome $Y$ is binary, where $Y=1$ (resp.\ 0) corresponds to a positive (resp.\ negative) outcome. This assumption will carry over to our incompatibility results in Section~\ref{section:incompatibility_results}. Note, however, that the proposed definitions can be amended to reflect (conditional) expectations or higher order moments when $Y \in \mathbb{R}$.

\begin{definition}[Statistical Parity in Outcomes]\label{def:statistical_parity_outcome} Statistical parity in outcomes is satisfied if the probabilities of experiencing a positive outcome are equal across groups, i.e.,
\begin{align*}
\mathbb{P}(Y(\mu(X))=1|G=g) = \mathbb{P}(Y(\mu(X))=1|G=g')  \\ \forall g, g' \in \sets{G}.
\end{align*}
\end{definition}

\begin{definition}[Conditional Statistical Parity in Outcomes]\label{def:conditional_statistical_parity_outcome} Conditional statistical parity in outcomes is satisfied if the probabilities of experiencing a positive outcome are equal across sensitive groups, conditional on some legitimate feature(s), i.e.,
\begin{align*}
\mathbb{P}(Y(\mu(X))=1|G=g, L=\ell) = \\ \mathbb{P}(Y(\mu(X))=1|G=g', L=\ell)  \\ \forall g, g' \in \sets{G}, \ell \in \sets{L}.
\end{align*}
\end{definition}

When the distribution of legitimate features vary widely between protected groups, we have discussed that conditional statistical parity is often preferable over statistical parity in the enrollment and allocation stages. This is not the case, however, when we want to ensure fairness in outcomes. We argue that statistical parity in outcomes is in fact the stronger condition for an allocation policy to satisfy. In the housing allocation problem, Black people are more often classified as being at higher risk to homelessness than White people. Enforcing conditional statistical parity merely protects the subpopulation in both groups who happen to have the same risk scores. However, we are more interested in seeing that both races \textit{overall} have an equal probability of experiencing a positive outcome, which may require allocating better/more resources to Black people because they are more at risk of homelessness in the first place. In general, our argument holds when there is existing and apparent discrimination between two or more groups, and ultimately our goal in enforcing statistical parity is to repair said discrimination.

The aforementioned definitions can be amended in multiple ways depending on the allocation setting. For instance, they can be altered so that one (historically marginalized) group has a higher probability of success than others. This change may be particularly helpful in settings involving affirmative action or reparations.

\section{Incompatibility Results}\label{section:incompatibility_results}

\subsection{Incompatibility of Allocation and Outcome Fairness}\label{subsection: incompatibility}

In this section, we present incompatibility results between fairness in allocation and fairness in outcomes by defining necessary conditions for different metrics to jointly hold and then providing intuitions as to why these conditions would not hold in general. Such results are important since policymakers may often want to see if they can achieve both allocation and outcome fairness between groups. Throughout this section, our results focus on the binary treatment case, i.e., $\mathcal{T} = \{0,1\}$, where $t=0$ corresponds to the no-treatment case (control). Each condition for the binary case can be extended to multiple treatments at the cost of complicating notation. While we do not explicitly consider budget constraints in deriving incompatibility results and in studying the fairness behavior of policies under various metrics, budget constraints are captured by the assignment probabilities. For example, $\mathbb{P}(\mu(X)=1)$, the population treatment probability, is bounded by the budget. For simplicity, we assume non-negative conditional average treatment effects, i.e., $\mathbb{E}[Y^1 - Y^0 \mid X=x] \geq 0 \; \forall x \in \mathcal{X}$, which in our motivating example means that giving someone (anybody) a housing resource is expected to help them. We discuss the incompatibility of allocation and outcomes in terms of the allocation rule $\mu$ and not the recommendation policy $\pi$ since the actual allocation is what affects the outcomes. Finally, our focus is on the case where policymakers can only influence outcomes through the allocation rule $\mu$ of limited resources and cannot improve the actual intervention effects. All proofs can be found in the Supplementary Materials.

First, we look at when we can expect statistical parity in allocation and statistical parity in outcomes to jointly hold. For a policy $\mu$ to satisfy statistical parity in allocation (Definition~\ref{def:statistical_parity_alloc}), it must be that group level assignment probabilities are equal across groups, $\mathbb{P}(\mu(X)=1|G=g) = \mathbb{P}(\mu(X)=1|G=g') \; \forall g, g' \in \mathcal{G}$, or equivalently, $\mathbb{P}(\mu(X)=1) = \mathbb{P}(\mu(X)=1|G=g) \; \forall g \in \mathcal{G}$. The following result shows that statistical parity in allocation and statistical parity in outcome will hold jointly only in exceptional circumstances.
\begin{proposition}
\label{prop: incompatible sp alloc and sp outcome} A policy $\mu$ satisfies both statistical parity in allocation (Definition~\ref{def:statistical_parity_alloc}) and statistical parity in outcomes (Definition~\ref{def:statistical_parity_outcome}) if and only if
\begin{equation}\label{eq: alloc outcome parity condition}
\begin{aligned}
&\textstyle  (\mathbb{E}[Y^1 - Y^0|G=g'] - \mathbb{E}[Y^1 - Y^0|G=g]) \mathbb{P}(\mu(X)=1)  \\
&= \mathbb{E}[Y^0 | G=g] - \mathbb{E}[Y^0|G=g'] \qquad \forall g,g' \in \mathcal{G}, g \neq g'.
\end{aligned}
\end{equation}    
\end{proposition}
Condition ~\eqref{eq: alloc outcome parity condition} requires setting policy $\mu$ so that the population probability of treatment assignment, when scaled by the treatment effect differences between $g$ and $g'$, is equal to the baseline outcome differences under no treatment between $g$ and $g'$. Since the probability of assignment is between~0 and~1, in order for ~\eqref{eq: alloc outcome parity condition} to hold, it must be that group~$g'$ has greater treatment effects than~$g$ in order to make up for group~$g$ having higher outcomes under no treatment. However, policymakers have no control over baseline outcomes and average treatment effects, and $g$ could have both better outcomes under no treatment and better treatment effects. Therefore, in general there may not exist a policy $\mu$ that satisfies \eqref{eq: alloc outcome parity condition}.

Since we may not be able to have statistical parity in both allocation and outcomes, perhaps we can have both conditional statistical parity (CSP) in allocation, where we condition on a set of legitimate features or risk scores, and statistical parity in outcomes. This is of practical interest since in our motivating example, under the U.S.\ Department of Housing and Urban Development's (HUD) guidance of prioritizing the most vulnerable populations \cite{hud}, scarce housing allocations are made based on risk scores measuring vulnerability under no treatment like the VI-SPDAT \cite{vispdatLA}. 

Let $\mathcal{L}_0$ denote the set of risk scores/levels that capture an individual's risk and $L_0$, which is just a function of the covariates as defined in Section~\ref{section:statistical_fainess}, denote a random variable that can take on values $\ell \in \mathcal{L}_0$. We assume $L_0$ is ``well chosen'' such that baseline outcomes under no treatment are independent of group membership for each score, i.e., $\mathbb{P}(Y^0 = 1| L_0 = \ell) = \mathbb{P}(Y^0 = 1|G= g, L_0=\ell) \; \forall g \in \mathcal{G}, \ell \in \mathcal{L}_0$. While this ``well chosen'' property may not hold for existing deployed risk scores, it is possible to construct risk scores that satisfy this property with well-calibrated probabilistic estimates. As in our motivating example where the risk scores are discrete, we also assume that $\mathcal{L}_0$ is a discrete set. We can then decompose group outcomes under no treatment for each group $g$ as $\mathbb{P}(Y^0=1|G=g) = \sum_{\ell \in \mathcal{L}_0} \mathbb{P}(Y^0=1|L_0=\ell) \mathbb{P}(L_0=\ell|G=g)$, which suggests group outcome differences are a result of group distribution differences across the risk scores in $\mathcal{L}_0$. It is natural then to enforce CSP in allocation conditioned on $L_0$ while trying to achieve outcome parity because we are accounting for distributional differences in vulnerability between groups. Let $\mu_{\mathcal{L}_0}$ be a policy that satisfies conditional statistical parity in allocation conditional on $L_0$. In particular, assume that $\mu_{\mathcal{L}_0}: \mathcal{L}_0 \to \mathcal{T}$ since in practice many policies are often simply a function of $L_0$ as described after Definition~\ref{def:conditional_statistical_parity_enroll}. 
 Also for ease of notation, let $\tau_{\ell,g} := \mathbb{E}[Y^1-Y^0|G=g, L_0 = \ell]$ be the average treatment effect for individuals in group $g$ with $L_0 = \ell$. We have the following incompatibility result.  
\begin{proposition}
\label{prop: csp baseline and sp outcome}
A policy $\mu_{\mathcal{L}_0}$ satisfies conditional statistical parity in allocation (Definition~\ref{def:conditional_statistical_parity_alloc}) and statistical parity in outcome (Definition~\ref{def:statistical_parity_outcome}) if and only if the following holds
\begin{equation}\label{eq: csp alloc baseline outcome parity condition}
\begin{split}
& \textstyle \sum_{\ell \in \mathcal{L}_0} \mathbb{P}(\mu_{\mathcal{L}_0}(\ell)=1) \left(\tau_{\ell,g} \mathbb{P}(L_0=\ell|G=g) \right. - \\ & \left. \qquad \tau_{\ell,g'} \mathbb{P}(L_0=\ell|G=g')\right)
 = \\
 & \mathbb{P}(Y^0=1|G=g') - \mathbb{P}(Y^0=1|G=g) \\ & \qquad \forall g,g' \in \mathcal{G}, g \neq g'.
\end{split}
\end{equation}
\end{proposition} 
We emphasize that Proposition \ref{prop: csp baseline and sp outcome} holds even if $L_0$ is not ``well-chosen'' since the proof does not depend on this property. What is interesting is that the result holds even if $L_0$ satisfies the desirable ``well-chosen'' property. Similar to Proposition \ref{prop: incompatible sp alloc and sp outcome}, there is no guarantee in general that there exists $\mu_{\mathcal{L}_0}$ satisfying ~\eqref{eq: csp alloc baseline outcome parity condition}. While policymakers can target specific risk groups through a $\mu_{\mathcal{L}_0}$ policy, they cannot control the treatment effects within each level, $\tau_{\ell,g}$. Therefore it might make sense to explicitly account for treatment effects as well. 

We propose another type of conditional statistical parity policy that uses an additional set of  legitimate features $L_1$ in addition to $L_0$, $\mu_{\mathcal{L}_0, \mathcal{L}_1}: \mathcal{L}_0 \times \mathcal{L}_1 \to \mathcal{T} $ , where $L_1$ is ``well chosen'' to capture an individual's treatment effects such that treatment effects are independent of $G$ and $X$ conditioned on $L_1$, i.e., $\mathbb{E}[Y^1 - Y^0| L_1 = l] = \mathbb{E}[Y^1 - Y^0|G= g,X=x, L_1 = l] \; \forall g \in \mathcal{G},  x \in \mathcal{X}, l \in \mathcal{L}_1$. Since the treatment effect can be captured by just $L_1$, for ease of notation we let $\tau_{l}: = \mathbb{E}[Y^1 - Y^0| L_1 = l]$ denote the treatment effect for individuals with treatment effects defined by $L_1 = l$. In practice, $L_1$ could be estimated via a heterogeneous treatment effects model, see e.g., \citet{wager2018estimation}. While $\mu_{\mathcal{L}_0}$ is commonly used in practice for allocation, we show that $\mu_{\mathcal{L}_0, \mathcal{L}_1}$ can mitigate disparities as well as, if not better than, $\mu_{\mathcal{L}_0}$. Intuitively this is because $\mu_{\mathcal{L}_0}$, as can be seen in~\eqref{eq: csp alloc baseline outcome parity condition}, does not account for treatment effects. By incorporating information about treatment effects, we can hope to further mitigate outcome disparities.
\begin{proposition} \label{prop: using two risk sets}
For any given conditional statistical parity policy $\mu_{\mathcal{L}_0}$, and under any distribution $\mathbb{P}$ over $(X, T, A, \{Y^t\}_{t\in \mathcal{T}})$, there exists $\mu_{\mathcal{L}_0, \mathcal{L}_1}$ such that 
\begin{equation}\label{eq: using two risks}
\begin{split}
& \textstyle \mathbb{P}(Y(\mu_{\mathcal{L}_0, \mathcal{L}_1}(L_0, L_1))=1|G=g) - \\ & \mathbb{P}(Y(\mu_{\mathcal{L}_0, \mathcal{L}_1}(L_0, L_1))=1|G=g')  \\
& \leq \;\;  \mathbb{P}(Y(\mu_{\mathcal{L}_0}(L_0))=1|G=g) - \\ & \mathbb{P}(Y(\mu_{\mathcal{L}_0}(L_0))=1|G=g') \;\; \forall g, g' \in \mathcal{G}, g \neq g'.
\end{split}
\end{equation}
Moreover, there exists a joint distribution $\mathbb{P}$ such that the inequality in ~\eqref{eq: using two risks} is strict for at least one pair g, g'.
\end{proposition}
Proposition~\ref{prop: using two risk sets} suggests that using more contextual information about group distributions or treatment effects may allow us to
define conditional statistical parity policies that are more fair in terms of outcomes. While we have explicitly defined $L_0$ for baseline outcomes and $L_1$ for treatment effects, we can simply define $L := \{(\ell,l) \mid \ell \in \mathcal{L}_0, l \in \mathcal{L}_1 \}$ and relabel each $(\ell,l)$ combination as some $\ell$ instead with the understanding that we are really partitioning our population by $L_0$ and $L_1$. In the remaining results, we use the combined $L$  and define $\tau_{\ell} = \mathbb{E}[Y^1 - Y^0| L= \ell]$. We also use $\mu_{\mathcal L}$ to denote a CSP in allocation policy conditioned on our new $L$. While using both $L_0$ and $L_1$ gives us more information and allocation flexibility, we show in the proposition below that incorporating both baseline outcome information and treatment effects may still not be enough to achieve statistical parity in outcomes across all groups. 
\begin{proposition} \label{prop: incompatible csp alloc and sp outcome}
An allocation policy $\mu_{\mathcal L}$ satisfies conditional statistical parity in allocation (Definition~\ref{def:conditional_statistical_parity_alloc}) and statistical parity in outcome (Definition~\ref{def:statistical_parity_outcome}) if and only if the following holds:
\begin{equation}\label{eq: csp alloc outcome parity condition tr eff}
 \begin{split}
  \textstyle \sum_{\ell \in \mathcal{L}} & \mathbb{P}(\mu_{\mathcal{L}}(\ell)=1) \tau_{\ell} \times \\
  [&\mathbb{P}(L=\ell|G=g)-
  \mathbb{P}(L=\ell|G=g')] = \\ & \mathbb{P}(Y^0=1|G=g') - \mathbb{P}(Y^0=1|G=g) \\ & \qquad \forall g, g' \in \mathcal{G}, g \neq g'. 
 \end{split}
 \end{equation}
\end{proposition} 
To gain some intuition on the implications of Proposition~\ref{prop: incompatible csp alloc and sp outcome} and when there exists $\mu_{\mathcal{L}}$ such that ~\eqref{eq: csp alloc outcome parity condition tr eff} will hold, let us look at a simple example. Suppose $|\mathcal{G}|=2, |\mathcal{L}|=2$, that is, there are only two groups and risk levels, $\mathbb{P}(Y^0=1|G=g) = 0.6$, and $\mathbb{P}(Y^0=1|G=g') = 0.4$. Let us also define $a_\ell:=\tau_{\ell} [\mathbb{P}(L=\ell|G=g)-\mathbb{P}(L=\ell|G=g')], \ell \in \{1,2\}$. Then in order for $\mu_{\mathcal{L}}$ to satisfy CSP in allocation conditional on $L$ and outcome parity, it follows from Proposition~\ref{prop: incompatible csp alloc and sp outcome} that
\begin{equation*}
\begin{split}
    & \mathbb{P}(\mu_{\mathcal{L}}(1) = 1) a_1 + \mathbb{P}(\mu_{\mathcal{L}}(2)=1) a_2 =\\
    & \mathbb{P}(Y^0=1|G=g')-\mathbb{P}(Y^0=1|G=g) = -0.2
\end{split}
\end{equation*}
if our allocation satisfies both CSP in allocation and outcome parity. Without loss of generality, let us assume $\mathbb{P}(L=1|G=g)>\mathbb{P}(L=1|G=g')$ and thus $\mathbb{P}(L=2|G=g)<\mathbb{P}(L=2|G=g')$, so that $a_1>0$ and $a_2<0$. We then have two possibilities: \emph{(i)} $a_2 \leq -0.2 \leq 0$ or \emph{(ii)} $-0.2<a_2$. In case (1), since $a_1$ and $a_2$ have opposite signs, we can easily find values $0<\mathbb{P}(\mu_{\mathcal{L}}(1)=1), \mathbb{P}(\mu_{\mathcal{L}}(2)=1)<1$ such that CSP in allocation and outcome parity are satisfied. However, in case (2), there does not exist such CSP allocation. The closest we could come to satisfying our conditions would be to set $\mathbb{P}(\mu_{\mathcal{L}}(1)=1)=0$ and $\mathbb{P}(\mu_{\mathcal{L}}(2)=1)=1$. From this example, we see that if the baseline difference in outcome between our two groups $|\mathbb{P}(Y^0=1|G=g)-\mathbb{P}(Y^0=1|G=g')|$ is very large, then it becomes harder to find an allocation that satisfies both CSP in allocation and outcome parity. This makes sense -- as the disparity in outcome under no treatment becomes very large, it becomes harder to make up this difference by allocation. 

\subsection{Policies with Outcome Fairness} 
In Section~\ref{subsection: incompatibility}, we established that in general there may not exist allocation policies that satisfy fairness in allocation (as measured by statistical and conditional statistical parity) and parity in outcomes. However, if a policymaker wants parity in outcomes, it is not immediately clear that a policy can even achieve this since outcomes depend on treatment effects and distribution of baseline risk within each group. In this section, we show that using feature information without group status, $X_{-G}$, in a policy can lead to policies with less outcome disparity than just using risk scores alone, but in general outcome parity may not be possible unless a policy explicitly uses group status. Let $\mu_{\mathcal{L},\mathcal{X}_{-G}}$ be a (possibly) randomized policy that uses $X_{-G}$ and $L$ for assignment. In the propositions that follow, for each inequality of outcome disparity between $g, g'$ under different policies, we use $g$ to denote the group with better outcomes under no treatment compared to those of $g'$. Therefore we do not have absolute values below since we only consider pairs of $g, g'$ where the outcome difference between $g$ and $g'$ is positive. Proposition~\ref{prop: using features better} shows that using extra information in the form of $X_{-G}$ leads to same, if not lower, disparities than policies relying only on $L$ across each comparison $ g, g' \in \mathcal{G}, g \neq g'$. 
\begin{proposition}\label{prop: using features better}
For any given policy $\mu_{\mathcal L}$, and under any distribution $\mathbb{P}$ over $(X, T, A, \{Y^t\}_{t\in \mathcal{T}})$, there exists $\mu_{\mathcal{L},\mathcal{X}_{-G}}$ such that 
\begin{equation}\label{eq: using features better}
 \begin{split}
 & \textstyle \mathbb{P}(Y(\mu_{\mathcal{L},\mathcal{X}_{-G}}(L, X_{-G}))=1|G=g) - \\ & \mathbb{P}(Y(\mu_{\mathcal{L},\mathcal{X}_{-G}}(L, X_{-G}))=1|G=g') \leq \\ &
 \mathbb{P}(Y(\mu_{\mathcal L}(L))=1|G=g) - \mathbb{P}(Y(\mu_{\mathcal L}(L))=1|G=g') 
 \end{split}
 \end{equation}
for all $g, g' \in \mathcal{G}, g \neq g'$ where $\mathbb{P}(Y^0=1|G=g) > \mathbb{P}(Y^0=1|G=g')$ and there exists a joint distribution $\mathbb{P}$ such that the inequality ~\eqref{eq: using features better} is strict for at least one pair g, g'.
\end{proposition}
The difficulty of reducing all pairwise outcome disparities lies in the fact that under any policy that does not explicitly use group status, outcomes for each group will have inter-dependencies. For example, we may reduce the outcome disparity between $g$ and $g'$ but also increase the outcome disparity between $g$ and $g''$ at the same time. Given this, we show below that a policy explicitly using group status will do better than a policy using all available features except for group status. Let $\mu_{\mathcal{L}, \mathcal{G}}$ denote a (possibly randomized) policy that uses $L$ and $G$ for assignment and emphasize that $\mu_{\mathcal{L}, \mathcal{G}}$ does not use the features $X$.
\begin{proposition}
\label{prop: using group better than features}
For any given policy $\mu_{\mathcal{L},\mathcal{X}_{-G}}$, and under any distribution $\mathbb{P}$ of $(X, T, A, \{Y^t\}_{t\in \mathcal{T}})$, there exists $\mu_{\mathcal{L}, \mathcal{G}}$ such that
\begin{equation}\label{eq: using group}
\begin{split}
&\textstyle \mathbb{P}(Y(\mu_{\mathcal{L}, \mathcal{G}}(L, G))=1|G=g) - \\ &\mathbb{P}(Y(\mu_{\mathcal{L}, \mathcal{G}}(L, G))=1|G=g') \leq \\
&\mathbb{P}(Y(\mu_{\mathcal{L},\mathcal{X}_{-G}}(L, X_{-G}))=1|G=g) - \\
& \mathbb{P}(Y(\mu_{\mathcal{L},\mathcal{X}_{-G}}(L, X_{-G}))=1|G=g')
\end{split}
\end{equation}
for all $g, g' \in \mathcal{G}, g \neq g'$ where $\mathbb{P}(Y^0)=1|G=g) > \mathbb{P}(Y^0=1|G=g')$ and there exists a joint distribution $\mathbb{P}$ such that the inequality ~\eqref{eq: using group} is strict for at least one pair g, g'. 
\end{proposition}
While the inequality in Proposition~\ref{prop: using group better than features} may not always be strict, one advantage of using group status is that if treatment effects are sufficient to reduce baseline outcome disparity, then there exists $\mu_{\mathcal{L}, \mathcal{G}}$ that can lead to outcome parity across all groups. To formalize this, let $g_{(0)}, g_{(1)}, \dots, g_{(|G|-1)}$ denote an ordering of our protected groups such that $\mathbb P(Y^0=1|G=g_{(0)}) \geq \mathbb P(Y^0=1|G=g_{(1)}) \geq \dots \geq \mathbb P(Y^0=1|G=g_{(|G| - 1)})$ so that the groups are ordered from highest to lowest of group outcome under no treatment. 
\begin{proposition}\label{prop: group disparity none}
If $\mathbb{E}[Y^1 - Y^0 | G=g_{(i)}] \geq \mathbb{P}(Y^0=1|G=g_{(0)}) - \mathbb{P}(Y^0=1|G=g_{(i)}) \; \forall i > 0$, then there exists $\mu_{\mathcal{L}, \mathcal{G}}$ such that 
\begin{equation}\label{eq: using group2}
 \begin{split}
 &\textstyle \mathbb{P}(Y(\mu_{\mathcal{L}, \mathcal{G}}(L, G))=1|G=g) - \\ & \mathbb{P}(Y(\mu_{\mathcal{L}, \mathcal{G}}(L, G))=1|G=g') = 0 \\
 & \forall g, g' \in \mathcal{G}, \quad  g \neq g'.
 \end{split}
 \end{equation}
\end{proposition}
Proposition~\ref{prop: group disparity none} implies that if we could treat all individuals in group $g_{(i)}, i \neq 0$, and the groupwise treatment effect was more than baseline differences under no treatment, then it is possible to allocate resources based on group status to achieve outcome parity.

\section{Discussion}\label{section:discussion}
In this paper we introduced a framework of measuring fairness in resource allocation systems at various stages of the allocation process. In settings like our motivating example, decision makers may desire a policy that is fair (e.g., imposing statistical parity) both in terms of allocation and outcomes. Our results show this is in general impossible. This incompatibility introduces tension between \textit{equality} (which comes from allocative fairness) and \textit{equity} (which comes from outcome fairness), leaving policymakers to have to choose between the two fairness goals. To add further nuance to this tension, we show that policies that are fair in allocation can result in better outcome fairness when that allocative fairness is achieved conditional on features measuring treatment effects rather than conditional on vulnerability. This provides a compromise of sorts where we could achieve allocative fairness while also inching toward better outcome fairness. However, doing so comes at the cost of not prioritizing the most vulnerable, which is often a desirable goal. Such ethical concerns need to be addressed by policymakers to determine the appropriate metrics of fairness evaluation.

We also show that using contextual information beyond baseline risk and treatment effects -- but without considering group status -- can lead to lower pairwise group disparities in terms of outcomes. On the other hand, explicit use of group status in addition to baseline risk and treatment effects can produce policies that are most outcome fair, which is often a desirable goal. However, using group status may pose practical and ethical issues since it is usually prohibited and it may be difficult to ensure such policies are not discriminatory even if well intended.

In the context of housing in Los Angeles, there is a desire on the part of community stakeholders to see fairness in all three stages: enrollment, allocation, and outcomes. As we have demonstrated, however, this desire is likely not possible.  Furthermore, while HUD seeks to prioritize based on vulnerability scores, we also show this does not guarantee equitable outcomes. Thus, policymakers will have to make difficult decisions as to which stage of the allocation process they value most to impose fairness constraints. It is also incumbent upon the research community to help community stakeholders understand why such decisions must be made (i.e., the logical impossibilities presented above) and the relative trade-offs in opting for fairness in different stages of the allocation process.

\section*{Acknowledgements}
P. Vayanos, B. Tang, S. Aghaei, and E. Rice gratefully acknowledge support from the Hilton C. Foundation, the Homeless Policy Research Institute, and the Home for Good foundation under the ``C.E.S. Triage Tool Research \& Refinement'' grant. P. Vayanos and K. Dullerud are also grateful for the support of the National Science Foundation under CAREER award number 2046230. B. Tang was also supported by the National Science Foundation Graduate Research Fellowship Program.


\bibliography{aaai23}

\end{document}